\documentclass[twocolumn,superscriptaddress,showpacs,amsmath,amssymb,prx]{revtex4-2}

\usepackage{graphicx}
\usepackage{color}
\usepackage{comment}
\usepackage[normalem]{ulem}
\usepackage{bm}

\begin{document}

\title{{Tuning the Parity Mixing of Singlet-Septet Pairing in a Half-Heusler Superconductor}} 

\author{K.~Ishihara}\email{ishihara@qpm.k.u-tokyo.ac.jp}
\author{T.~Takenaka}
\author{Y.~Miao}
\author{Y.~Mizukami}
\author{K.~Hashimoto}
\affiliation{Department of Advanced Materials Science, University of Tokyo, Kashiwa, Chiba 277-8561, Japan}

\author{M.~Yamashita}
\affiliation{The Institute for Solid State Physics, University of Tokyo, Kashiwa, Chiba 277-8581, Japan}

\author{M.~Konczykowski}
\affiliation{Laboratoire des Solides Irradi{\' e}s, CEA/DRF/IRAMIS, Ecole Polytechnique, CNRS, Institut Polytechnique de Paris, F-91128 Palaiseau, France}

\author{R.~Masuki}
\affiliation{Department of Applied Physics, University of Tokyo, Hongo, Tokyo 113-8656, Japan}
\author{M.~Hirayama} 
\affiliation{RIKEN Center for Emergent Matter Science (CEMS), Wako 351-0198, Japan}
\author{T.~Nomoto}
\affiliation{Department of Applied Physics, University of Tokyo, Hongo, Tokyo 113-8656, Japan}
\author{R.~Arita} 
\affiliation{Department of Applied Physics, University of Tokyo, Hongo, Tokyo 113-8656, Japan}
\affiliation{RIKEN Center for Emergent Matter Science (CEMS), Wako 351-0198, Japan}

\author{O.~Pavlosiuk}
\author{P.~Wi{\' s}niewski}
\affiliation{Institute of Low Temperature and Structure Research, Polish Academy of Sciences, Ok\'{o}lna 2, 50-422 Wroc{\l}aw, Poland}
\author{D.~Kaczorowski}
\affiliation{Institute of Low Temperature and Structure Research, Polish Academy of Sciences, Ok\'{o}lna 2, 50-422 Wroc{\l}aw, Poland}
\affiliation{Institute of Molecular Physics, Polish Academy of Sciences, Smoluchowskiego 17, 60-179 Pozna{\' n}, Poland}

\author{T.~Shibauchi}\email{shibauchi@k.u-tokyo.ac.jp}
\affiliation{Department of Advanced Materials Science, University of Tokyo, Kashiwa, Chiba 277-8561, Japan}


\begin{abstract}
In superconductors, electrons with spin ${s=1/2}$ form Cooper pairs whose spin structure is usually singlet (${S=0}$) or triplet (${S=1}$). When the electronic structure near the Fermi level is characterized by fermions with angular momentum ${j=3/2}$ due to strong spin-orbit interactions, novel pairing states such as even-parity quintet (${J=2}$) and odd-parity septet (${J=3}$) states are allowed. Prime candidates for such exotic states are half-Heusler superconductors, which exhibit unconventional superconducting properties, but their pairing nature remains unsettled. Here we show that the superconductivity in the noncentrosymmetric half-Heusler LuPdBi can be consistently described by the admixture of isotropic even-parity singlet and anisotropic odd-parity septet pairing, whose ratio can be tuned by electron irradiation. From magnetotransport and penetration depth measurements, we find that carrier concentrations and impurity scattering both increase with irradiation, resulting in a nonmonotonic change of the superconducting gap structure. Our findings shed new light on our fundamental understanding of unconventional superconducting states in topological materials.
\end{abstract}

\maketitle


\section{Introduction}

Half-Heusler materials, $R$PtBi and $R$PdBi, where $R$ is a rare-earth element, are of notable interest as they can provide a new platform for topological phenomena~\cite{Chadov2010,Lin2010,Xiao2010}. Their outstanding feature is that the strength of the spin-orbit interaction (SOI) is controllable by changing the constituent elements. When the SOI is strong enough, the $s$-like conduction band is pushed down, and near the Fermi level, the $\Gamma_8$ band having fourfold degeneracy at the $\Gamma$ point forms a $p$-like band with total angular momentum $j=3/2$. Related to this strong SOI, several nontrivial topological phenomena, such as the Dirac surface state~\cite{Liu2016}, the anomalous Hall effect~\cite{Suzuki2016}, chiral anomaly~\cite{Hirschberger2016}, and the planar Hall effect~\cite{Kumar2018}, have been reported experimentally.

\begin{figure}[t]
    \includegraphics[width=\linewidth]{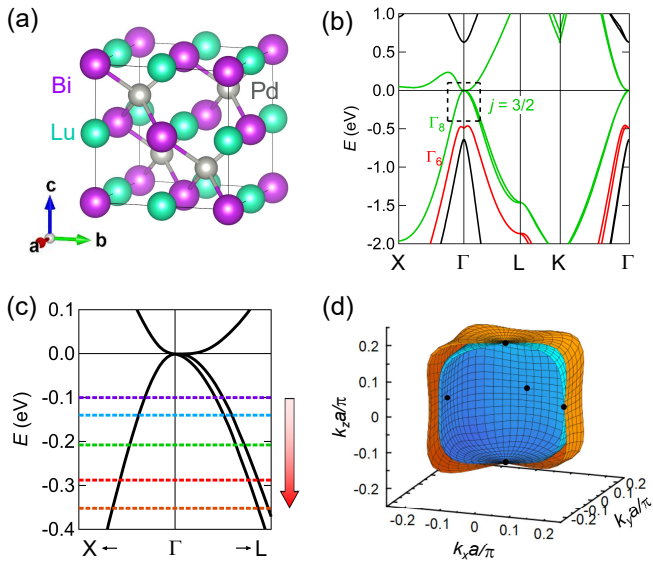}
    \caption{{Crystal and electronic structures of LuPdBi.} {(a)} Crystal structure of the half-Heusler superconductor LuPdBi.   {(b)} {\it Ab initio} calculations of the band structure near the Fermi energy $E_{\rm F}$. The green and red lines are $\Gamma_8$ and $\Gamma_6$ bands, respectively.  {(c) } Enlarged view of the calculated band structure near the $\Gamma$ point [dashed square in (b)]. The horizontal purple, light-blue, green, red, and brown dashed lines represent chemical potentials obtained from the carrier densities in the pristine, 3.13, 5.28, 8.44, and 11.61\,C/cm$^2$ irradiated samples, respectively. The red arrow illustrates the shift of the chemical potential caused by electron irradiation. 
    {(d)} Fermi surfaces calculated from the ${\bf k} \cdot {\bf p}$ Hamiltonian [Eq.\,(\ref{kp_model})] with chemical potential $u = 0.1$\,eV. Orange and blue surfaces are the outer and inner spin-split FSs, respectively. Black dots represent the points where the split size of two FSs becomes zero. }
    \label{structure}
\end{figure}

Recently, superconductivity has been reported in half-Heusler materials~\cite{Butch2011,Pavlosiuk_LuPdBi,Nakajima2015,Meinert2016,Kim2018}, even though they have very low carrier densities, typically of the order of 10$^{18}$--10$^{19}$\,cm$^{-3}$. Since the crystal structure of half-Heusler materials has no inversion center (Fig.\,\ref{structure}(a)), they are classified as noncentrosymmetric superconductors having spin-split Fermi surfaces (FSs). In such superconductors with no inversion center, even-parity and odd-parity pairing states are allowed to admix~\cite{Gor'kov2001,Frigeri2004,Bauer2012}, which has been experimentally shown by the contrasting superconducting gap structures in Li$_2$Pd$_3$B and Li$_2$Pt$_3$B with different SOI~\cite{Yuan2006,Nishiyama2007}. Furthermore, in half-Heusler superconductors, $j=3/2$ fermions form a pairing bound state, and thus Cooper pairs with not only standard even-parity singlet ($J=0$) and odd-parity triplet ($J=1$) states, but also even-parity quintet ($J=2$) and odd-parity septet ($J=3$) states~\cite{Brydon2016,Dutta2021}. While experimental studies on YPtBi have reported spin-split FSs and a superconducting gap structure with line nodes~\cite{Kim2018,Kim_arxiv}, the pairing symmetries of half-Heusler superconductors are still under debate~\cite{Brydon2016,Boettcher2018,Venderbos2018,Yu2018,Roy2019,Sim2019}.

The possible on-site pairing states in the $T_d$ point group can be classified into an $s$-wave $A_1$ singlet state or $d$-wave $E$ and $T_2$ quintet states~\cite{Brydon2016}. In the $A_1$ pairing state, the inversion symmetry breaking leads to an admixture of $s$-wave singlet and $p$-wave septet states. When the odd-parity $p$-wave component is sufficiently large, the superconducting gap structure becomes very anisotropic having line nodes. In this case, the line nodes are not protected by the symmetry, and their positions can be shifted by the ratio of the even- and odd-parity components. Here, for the odd-parity component, the $p$-wave septet pairing is more favorable than the triplet one, which is expected to have a higher-momentum $f$-wave symmetry~\cite{Brydon2016}. In the $E$ and $T_2$ pairing states, time-reversal symmetry-breaking states with a nodal gap structure become energetically stable. In this case, the nodes are protected by the $d$-wave symmetry of the order parameter. In previous studies~\cite{Kim2018}, however, it has not been clarified whether the nodes in the superconducting gap are symmetry protected or not. We note that such a nodal singlet-septet state and time-reversal symmetry-breaking $E$ or $T_2$ states can have different exotic topological properties~\cite{Timm2017,Wang2017,Kobayashi2019,Roy2019}. Moreover, recent theoretical studies have suggested other types of possible topological superconducting states with $s$-wave singlet and $d$-wave quintet components~\cite{Yu2018,Sim2019}. Therefore, it is important to clarify the pairing state in half-Heusler superconductors for both fundamental and applied research.

In mixed-parity states, the amplitude of the odd-parity superconducting gap component is proportional to the antisymmetric SOI (ASOI) vector ${\bm g_k}$ relevant for the superconducting pairing~\cite{Frigeri2004}. Here $|{\bm g_k}|$ is proportional to $|{\bm k}|$ in the low-energy limit. This property indicates that if the Fermi wave number $k_{\rm F}$ is increased by carrier doping, the odd-parity gap component is enhanced, which makes the superconducting gap more anisotropic. Each of the two split FSs has the odd-parity ($p$-wave) anisotropic gap with the opposite sign, in addition to the even-parity ($s$-wave) isotropic gap with the same sign. Thus, the superconducting state may exhibit a multigap behavior when the anisotropic component is large enough. For instance, the gap on one FS has nodes at certain ${\bm k}$ points, and the other FS is fully gapped. On the other hand, when nodes are protected by superconducting symmetry as in the $E$ or $T_2$ pairing states, the carrier doping has little influence on the gap anisotropy. Thus, in this study, we focus on the carrier doping effects and the multigap properties to discuss the pairing state of the half-Heusler superconductor LuPdBi with the highest superconducting transition temperature $T_{\rm c}\approx1.8$\,K among half-Heusler superconductors~\cite{Pavlosiuk_LuPdBi,
Nakajima2015}.

We use high-energy electron irradiation to tune carrier concentrations in a controlled way by introducing nonmagnetic Frenkel pairs, which are known to act as donors or acceptors that shift the chemical potential in materials with low carrier density~\cite{Billington1961,Zhao2016}. On the other hand, it is expected that the Frenkel pairs also act as nonmagnetic scatterers~\cite{Mizukami2014,Cho2016,Cho2018,Mizukami2017,Takenaka2017,Ishihara2018}, and the introduced impurity scattering makes the superconducting gap more isotropic~\cite{Mizukami2014,Cho2018}. Therefore, we need to consider two conflicting effects of irradiation on the superconducting gap structure, carrier doping and impurity scattering effects, which can be evaluated independently from magnetotransport measurements.
We find from magnetic penetration depth measurements that the gap structure in LuPdBi changes from nodeless to nodal, and then to nodeless again, by increasing the irradiation dose, which can be semiquantitatively explained by the singlet-septet pairing state taking into account the above two effects.

\section{Methods}

Single crystals of LuPdBi were grown from Bi-flux using elemental constituents in a 1:1:12 atomic ratio as described in detail in Ref.~\cite{Pavlosiuk_LuPdBi}. 
The crystals were found homogeneous and free of foreign phases, with the chemical compositions very close to the ideal one (see Appendix B). 
The excellent quality of the crystals was also confirmed by observation of Shubnikov-de Haas oscillations in temperatures up to 10\,K~\cite{Pavlosiuk_LuPdBi}. 
For the magnetic penetration depth measurements, the samples were cut into a platelike shape with typical dimensions of $350 \times 350 \times 50\,\mu$m$^3$. 


We have calculated the band structure of LuPdBi using the WIEN2k implementation~\cite{WIEN2k} of the full-potential linearized augmented plane-wave method with the modified Becke-Johnson exchange-correlation potential~\cite{Tran2009}, including the spin-orbit coupling. The muffin-tin radii ($R_{\rm MT}$) of 2.5 bohr was used, and the maximum modulus for the reciprocal vectors $K_{\rm max}$ was chosen such that $R_{\rm MT}K_{\rm max}=7.0$. The $k$-point mesh was taken to be $10 \times 10 \times 10$ in the first Brillouin zone. The cubic lattice constant $a$ was set to be 657\,pm.


Electron irradiation was done by using the SIRIUS Pelletron linear accelerator operated by the Laboratoire des Solides Irradi{\'e}s (LSI) at {\'E}cole Polytechnique, with an incident electron energy of 2.5\,MeV~\cite{Mizukami2014}. The samples were kept at about 20\,K in a liquid hydrogen bath during the irradiation to prevent defect migration and agglomeration. Since the penetration range of the irradiated electrons in LuPdBi was about 1.8\,mm which is much longer than the sample thickness, the point defects were introduced homogeneously. Partial annealing of the introduced defects occured upon warming to room temperature and during the sample transfer. 


The temperature dependence of zero-field resistivity was measured by the standard four-probe method in a $^3$He refrigerator with an ac current of less than 100\,$\mu$A. The electrical contacts were made on the surface by using silver paste. The magnetoresistance and Hall effect down to 2\,K were measured using a Physical Property Measurement System (PPMS) from Quantum Design in magnetic fields up to 9\,T. The applied ac current was 2\,mA. The longitudinal resistivity was measured by the standard four-probe or van der Pauw methods, and the Hall resistivity was measured by attaching the electrical contacts to make the voltage wirings perpendicular to the current wirings. For the $H_{\rm c2}$ measurements, we used the standard four-probe method and a dilution refrigerator that can reach down to about 100\,mK. The applied ac current was reduced to less than 3.16\,$\mu$A to avoid Joule heating. The irradiated samples for the $H_{\rm c2}$ measurements were cut from the same crystal with the pristine crystal (\#1). 


The temperature dependence of the change in the magnetic penetration depth $\Delta \lambda (T) \equiv \lambda (T) - \lambda (0)$ was measured down to 40\,mK using a tunnel diode oscillator technique with a resonant frequency of about 13.8\,MHz in a dilution refrigerator. 
The frequency shift was given by $\Delta f (T) = -(f_0 V_{\rm s} / 2V_{\rm c} (1-N)) \Delta \chi (T)$, where $f_0$ is the resonant frequency without the sample, $V_{\rm s}$ and $V_{\rm c}$ are the sample and coil volumes, respectively, $N$ is the demagnetization factor, and $\Delta \chi (T)$ is the shift of the magnetic susceptibility in the SI base units. With the characteristic sample size $R$, $\chi = (\lambda / R) {\rm tanh} (R / \lambda) - 1$, from which we can calculate the $\Delta \lambda (T)$~\cite{Prozorov2000}. 
For the penetration depth measurements, we used different crystals for the pristine, 3.13, and 8.44\,C/cm$^2$ irradiated samples, and the 3.13\,C/cm$^2$ (8.44\,C/cm$^2$) irradiated sample was subsequently irradiated to 5.28\,C/cm$^2$ (11.61\,C/cm$^2$) after the measurements.

\section{Results}
\subsection{Electronic structure}
Our {\it {ab initio}} calculations of the electronic structure reveal that the $\Gamma_8$ band formed by $j=3/2$ fermions dominates near the Fermi energy $E_{\rm F}$ [Fig.\,\ref{structure}(b)], which confirms that LuPdBi gives a platform of the $j=3/2$ superconductivity. To model the $\Gamma_8$ band, we use a low-energy $j=3/2$ ${\bf k} \cdot {\bf p}$ theory \cite{Brydon2016,Kim2018}, and the effective Hamiltonian can be written as
\begin{eqnarray}
H =& \alpha k^2 + \beta \sum_i k_i^2 {\breve J}_i^2 + \gamma \sum_{i \neq j} k_i k_j {\breve J}_i {\breve J}_j \nonumber\\ 
&+ \delta \sum_i k_i \left( {\breve J}_{i+1} {\breve J}_i {\breve J}_{i+1} - {\breve J}_{i+2} {\breve J}_i {\breve J}_{i+2} \right) ,
\label{kp_model}
\end{eqnarray}
where $i=x$, $y$, $z$ and $i+1=y$ for $i=x$ etc., and ${\breve J}_i$ are the matrix representations of the $j=3/2$ angular momentum operators. This effective Hamiltonian reasonably reproduces the low-energy results of {\it ab initio} calculations 
(see Appendix A), confirming that the effective $k$-linear relation of the ASOI term is applicable. In real materials, impurities inevitably present cause a chemical potential shift [Fig.\,\ref{structure}(c)], and an example of the spin-split FSs with the chemical potential of $-0.1$\,eV is shown in Fig.\,\ref{structure}(d). The split size of FSs is zero for the [100] and its equivalent directions [black points in Fig.\,\ref{structure}(d)], and largest for the [111] and its equivalent directions.

\begin{figure}[t]
    \includegraphics[width=\linewidth]{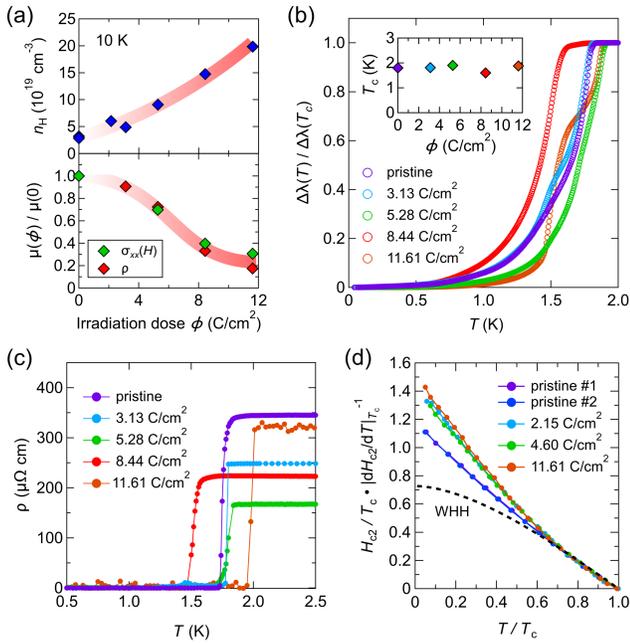}
    \caption{{Physical properties of pristine and electron-irradiated LuPdBi.} (a) Upper and lower panels depict the carrier density $n_{\rm H}$ (blue diamonds) and normalized mobility $\mu(\phi)/\mu(0)$ (green and red diamonds) extracted from magnetotransport measurements at $T=10$\,K as a function of the irradiation dose $\phi$, respectively. The green and red diamonds represent the $\mu(\phi)/\mu(0)$ values calculated from the magnetoconductance $\sigma_{xx} (H)$ and resistivity $\rho$, respectively (see Appendix D).  (b) Normalized change of penetration depth as a function of temperature $\Delta \lambda (T)$ in the pristine (purple), 3.13 (light-blue), 5.28 (green), 8.44 (red), and 11.61\,C/cm$^2$ (brown) irradiated samples. Inset shows $T_{\rm c}$ [defined at the onset of $\Delta \lambda (T)$] as a function of irradiation dose. (c) Temperature dependence of resistivity at low temperatures in the pristine and irradiated samples.  (d) Reduced upper critical field as a function of $T/T_{\rm c}$ in the pristine (purple and blue), 2.15 (light-blue), 4.80 (green), and 11.61\,C/cm$^2$ (brown) irradiated samples. The black broken line represents the standard temperature dependence of $H_{\rm c2}$ in the WHH model.}
    \label{irradiation}
\end{figure}

The carrier density $n_{\rm H}$ and the normalized carrier mobility $\mu(\phi)/\mu(0)$ as a function of the electron irradiation dose $\phi$ estimated from magnetotransport measurements (see Appendix D) in pristine and electron-irradiated LuPdBi samples are shown in Fig.\,\ref{irradiation}(a). The positive Hall coefficient and the increase of $n_{\rm H}$ indicate that the irradiation shifts the chemical potential downward [Fig.\,\ref{structure}(c)], making the volume of the hole FS as well as $k_{\rm F}$ larger. The rigid band picture is supported by the small defect density, estimated to be of the order of 0.001 d.p.a.\ (displacements per atom) in the most irradiated crystal in this study. We also note that the FS volume is always less than 1\% of the Brillouin zone, which implies that the low-energy ${\bf k} \cdot {\bf p}$ theory can be applied to the present system.

\subsection{Superconducting transition and upper critical field}

The change in the magnetic penetration depth $\Delta \lambda (T)$ normalized by the value at $T_{\rm c}$ is shown in Fig.\,\ref{irradiation}(b) for pristine and electron-irradiated samples. The superconducting transition temperatures $T_{\rm c}$ of irradiated samples determined by the onset of $\Delta \lambda (T)$ fall within $\pm 15$\,\% of that of the pristine one, which is consistent with the resistivity measurements [Fig.\,\ref{irradiation}(c)]. 
Near $T_{\rm c}$, the penetration depth is expected to become comparable to the sample size because $\lambda (0)$ in the low-carrier half-Heusler superconductors is several $\mu$m long~\cite{Bay2014,Pavlosiuk_LuPdBi}. Owing to this size-limiting effect, the shape irregularity and small inhomogeneity inevitably present in the sample tend to make the transitions near $T_{\rm c}$ relatively broad compared with the resistive transitions~\cite{Kim2018}, and sometimes a shoulder-like structure can be found in $\Delta \lambda (T)$. At sufficiently low temperatures $T\ll T_{\rm c}$, however, such a size-limiting effect is not an issue, and thus we focus on the low-temperature $\Delta \lambda (T)$ in the next section to discuss the gap structure in LuPdBi.

Figure\,\ref{irradiation}(d) shows the reduced upper critical field normalized by the initial slope, $H_{\rm c2}/T_{\rm c}\cdot |dH_{\rm c2}  / dT|_{T_{\rm c}}^{-1}$, as a function of reduced temperature, $T/T_{\rm c}$, in the pristine and irradiated samples. We find two characteristic features in Fig.\,\ref{irradiation}(d). First, the normalized upper critical fields in all samples show an upward curvature in the whole $T$ range, which is in contrast with the conventional Werthamer-Helfand-Hohenberg (WHH) temperature dependence. 
There are several possible origins giving rise to an upward curvature in $H_{\rm c2}(T)$, such as multigap effects~\cite{Gurevich2003}, strong coupling effects~\cite{Thomas1996}, the coexistence of superconductivity and ferromagnetism~\cite{Aoki2019}, 
and critical spin fluctuations \cite{Tada2008}. Considering that LuPdBi is a nonmagnetic material with split FSs, the effect of the multiband is the most plausible to explain the upward curvature in the present case. Second, at the lowest temperatures, $H_{\rm c2}$ values for some samples exceed the Pauli limit estimated from the simple BCS theory (see Appendix E). This high $H_{\rm c2}$ is consistent with the existence of the odd-parity pairing component, but other sources may enhance the Pauli limit from the BCS value, including strong coupling superconductivity and spin-orbit scattering. Thus this alone cannot be taken as decisive evidence for the odd-parity pairing in LuPdBi.

\subsection{Irradiation effects on superconducting gap structure}

\begin{figure*}[t]
    \includegraphics[width=\linewidth]{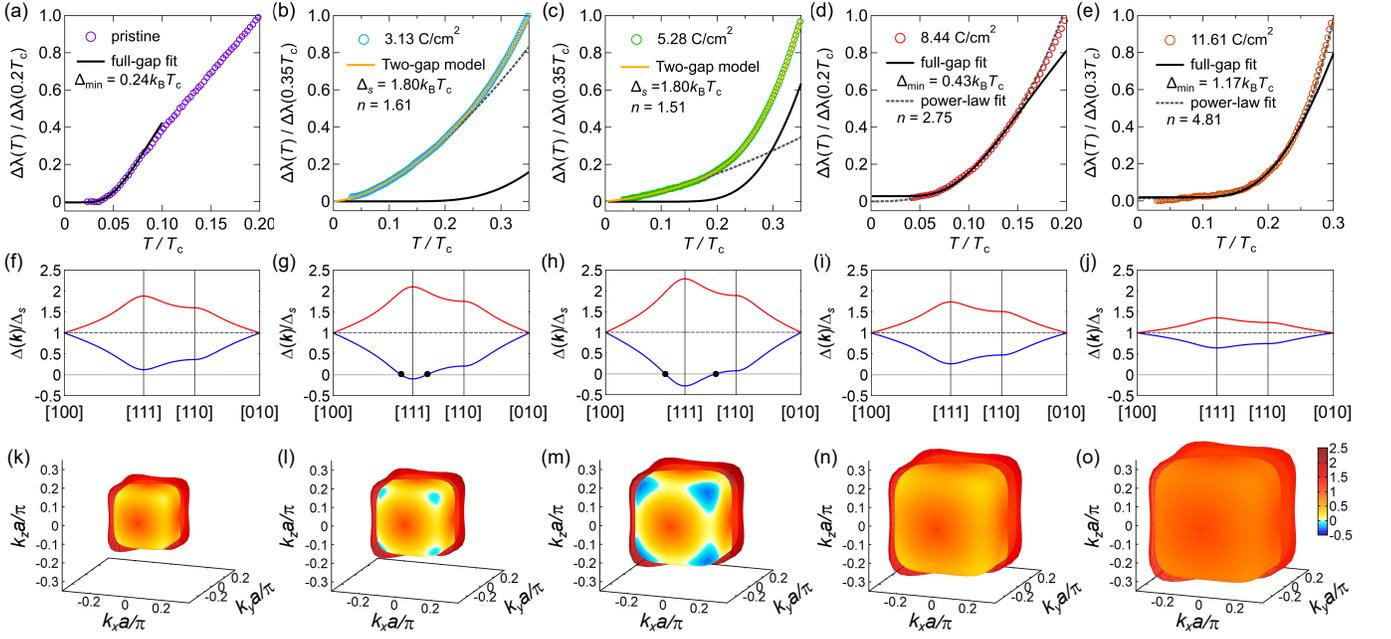}
    \caption{{Evolution of penetration depth and superconducting gap structure with electron irradiation.} (a)-(e) $\Delta \lambda (T)$ normalized by the value at $0.2$\,$T_{\rm c}$ or $0.35$\,$T_{\rm c}$ for ({a}) pristine, ({b}) 3.13, ({c}) 5.28, ({d}) 8.44, and (e) 11.61\,C/cm$^2$ irradiated samples. Open circles are experimental data, and black solid and dashed lines in panels (a), (d), and (e) are fitting curves in the fully gapped and power-law models, respectively. Black solid and broken lines in panels (b) and (c) are the fully gapped and power-law components of the two-gap model described by Eq.\,(\ref{fit}), respectively.  (f)-(j) Momentum direction dependence of the normalized superconducting gap in the pristine and irradiated samples derived from the effective ${\bf k} \cdot {\bf p}$ model in Eq.\,(\ref{kp_model}) and $\Delta_{p\rm {,max}} / \Delta_s$ values calculated from the simple model described by Eq.\,(\ref{Delta_p}). Red and blue lines correspond to the gap size on the outer and inner FSs, respectively. Black circles represent the positions of nodes. (k)-(o) Corresponding three-dimensional superconducting gap structure in the pristine and irradiated samples. White lines in panels (l) and (m) represent the nodal lines.}
    \label{gap}
\end{figure*}

Next, we discuss $\Delta \lambda(T)$ at low temperatures, which directly reflects the superconducting gap structure. In the case of the pristine sample, $\Delta \lambda (T)$ between $0.05$\,$T_{\rm c}$ and $0.2$\,$T_{\rm c}$ shows quasilinear $T$ dependence, while below about 0.07\,$T_{\rm c}$ it deviates from the linear behavior and shows a saturation [Fig.\,\ref{gap}(a)] that can be approximated as about $T^{3.5}$. This saturating behavior with a high exponent indicates a fully gapped state with small gap minima. By fitting the data below $0.07$\,$T_{\rm c}$ with a fully gapped model, $\Delta \lambda \propto T^{-1/2} \exp(- \Delta_{\rm{min}} / k_{\rm B} T)$, where $\Delta_{\rm{min}}$ is the minimum gap and $k_{\rm B}$ is the Boltzmann constant, we obtain $\Delta_{\rm{min}} = 0.24$\,$k_{\rm B} T_{\rm c}$, which is much smaller than the BCS value of 1.76\,$k_{\rm B} T_{\rm c}$. We note that between about 0.07\,$T_{\rm c}$ and 0.13\,$T_{\rm c}$, we find sublinear temperature dependence of $\Delta \lambda(T)$, which is similar to the case when the system is close to transitions between different types of gap-node topology~\cite{Mazidian2013,Cho2016}.  Therefore, we conclude that a highly anisotropic fully gapped state with deep gap minima is realized in the pristine sample. This result rules out a symmetry-protected nodal gap structure, which has been proposed for even-parity ($d$-wave) quintet states with time-reversal symmetry breaking~\cite{Brydon2016}. 
Note that the sub-linear dependence has not been found in the $\Delta \lambda(T)$ measurements down to $\sim 0.06T_{\rm c}$ for YPtBi~\cite{Kim2018}. This supports the nodal state in YPtBi, which can be accounted for by a more anisotropic superconducting gap associated with larger SOI than those in LuPdBi.

The low-$T$ behavior of $\Delta \lambda (T)$ is significantly changed by the electron irradiation. In the 3.13 and 5.28\,C/cm$^2$ irradiated samples, $\Delta \lambda (T)$ does not show a saturation down to the lowest $T$, and it follows power-law $T^n$ dependence below $0.15$\,$T_{\rm c}$ with the exponent $n \leq 2$, indicative of a nodal gap structure. We have checked that this conclusion is robust against the choice of maximum fitting temperature up to about 0.20\,$T_{\rm c}$, above which the power-law fit no longer reproduces the data. In the singlet-septet model, even if one of the split FSs has a nodal gap structure, the other FS keeps a fully gapped state. Thus we fit the experimental data at low temperatures using a two-gap model given by
\begin{equation}
\Delta \lambda (T) = A T^{-1/2} \exp \left( -\frac{\Delta_{s}}{k_{\rm B} T} \right) + B T^n ,
\label{fit}
\end{equation}
where $A$ and $B$ represent the weights of the two contributions, $\Delta_{s}$ is the excitation gap of fully-gapped one, and the exponent $n$ is expected between 1 and 2 for line nodes. By fixing $\Delta_{s} = 1.80$\,$k_{\rm B} T_{\rm c}$ close to the BCS value, we can reproduce the experimental data up to $0.3$\,$T_{\rm c}$ for the 3.13 and 5.28\,C/cm$^2$ samples, with $n=1.61$ and 1.51, respectively [Figs.\,\ref{gap}(b) and (c)]. 

In the 8.44\,C/cm$^2$ irradiated sample, the power-law fitting, $\Delta \lambda (T) \propto T^n$, below $0.15$\,$T_{\rm c}$ gives an exponent of $n = 2.75$. For the 11.61\,C/cm$^2$ irradiated sample, the low-temperature $\Delta \lambda (T)$ shows a flat behavior, which gives a higher exponent $n=4.81$ in the power-law analysis. Such high-power temperature dependence is not expected from the gap with line nodes, and experimentally indistinguishable from the exponential dependence (Figs.\,\ref{gap}(d) and (e)). This indicates the reemergence of a fully-gapped state in the high dose region. By fitting the data using a fully-gapped model, the minimum gap $\Delta_{\rm {min}}$ is estimated as $0.43$\,$k_{\rm B} T_{\rm c}$ ($1.17$\,$k_{\rm B} T_{\rm c}$) for the 8.44\,C/cm$^2$ (11.61\,C/cm$^2$) irradiated sample.

\begin{figure}[t]
    \includegraphics[width=0.85\linewidth]{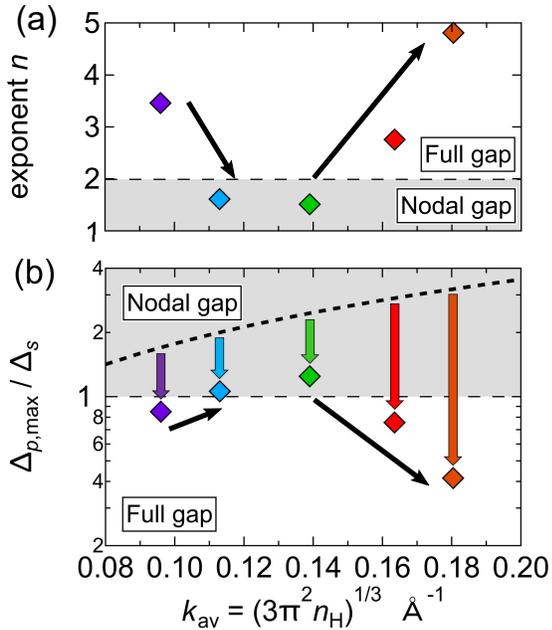}
    \caption{{Doping dependence of the penetration depth exponent and the septet gap component.} (a) Exponent $n$ in the power-law $T^n$ dependence of $\Delta \lambda (T)$ as a function of the effective Fermi wave vector $k_{\rm av} = (3\pi^2 n_{\rm H})^{1/3}$ for the pristine (purple), 3.13 (light-blue), 5.28 (green), and 8.44 (red), and 11.61 (brown) C/cm$^2$ irradiated samples. The exponent $n \leq 2$ (shaded region) indicates a nodal gap structure, while $n >2$ implies a fully gapped state.  (b) Ratio of the $p$-wave septet component $|\Delta_{p{\rm ,max}}|$ to the $s$-wave component $|\Delta_s|$ as a function of $k_{\rm av}$. The black dashed line represents the relation $\Delta_{p{\rm ,max}} \propto k_{\rm av}$. The lengths of the colored vertical arrows are proportional to $1/\mu$ values [see Eq.\,(\ref{Delta_p})] in each sample. In the shaded region, a nodal gap structure is expected. 
    }
    \label{exponent}
\end{figure}

The irradiation evolution of the exponent $n$ obtained from the power-law fitting of low-temperature $\Delta \lambda (T)$ is summarized in Fig.\,\ref{exponent}(a). The exponent less than 2 implies that the gap function has line nodes, where the deviation from $n=1$ toward $n=2$ is expected due to impurity scattering~\cite{Mizukami2014}. The nonmonotonic change between the fully gapped and nodal states with the irradiation dose immediately indicates that the line nodes are not protected by symmetry. This is consistent with the singlet-septet pairing gap functions. In the following, we will make a more quantitative analysis based on the singlet-septet model.

\section{Discussion}

As mentioned in the introduction, the amplitude of the $p$-wave septet component is proportional to $k_{\rm F}$. Considering that the FSs in LuPdBi are nearly isotropic owing to the low carrier concentrations, we simply assume $k_{\rm {av}} = (3\pi^2 n_{\rm H})^{1/3}$ as the averaged Fermi wave number. On the other hand, impurity scattering makes the gap more isotropic and thus reduces the anisotropic septet component. Theoretical studies on the impurity effect of anisotropic superconducting gap have shown that the gap minima increase almost linearly with the impurity scattering rate $1/\tau$~\cite{Mishra2009}. Thus we assume that the size of the reduced septet component is proportional to the scattering rate $1/\tau \propto 1/\mu$. Then the amplitude of the septet component, $\Delta_{p{\rm ,max}}$, can be represented as
\begin{equation}
\Delta_{p{\rm {,max}}} = Ck_{\rm {av}} - \frac{D}{\mu},
\label{Delta_p}
\end{equation}
where $C$ and $D$ are positive constants. From the full-gap analysis for the pristine, 8.44, and 11.61\,C/cm$^2$ irradiated samples (Figs.\,\ref{gap}(a), (d), and (e)), the obtained values of $\Delta_{\rm {min}} = \Delta_s-\Delta_{p\rm {,max}}$ are used to estimate $\Delta_{p\rm {,max}} / \Delta_s \approx 0.87$, 0.76, and 0.35, respectively, with the isotropic component $\Delta_s=1.80$\,$k_{\rm B} T_{\rm c}$. The coefficients $C$ and $D$ in Eq.\,(\ref{Delta_p}) are determined by the fitting of the above $\Delta_{p\rm {,max}} / \Delta_s$ values in the three samples. From this analysis based on Eq.\,(\ref{Delta_p}), we can estimate $\Delta_{p\rm {,max}} / \Delta_s=1.10$ and 1.29 for the 3.13 and 5.28\,C/cm$^2$ samples, respectively (Fig.\,\ref{exponent}(b)). These values of anisotropic/isotropic gap ratio greater than unity signify the existence of line nodes, which is consistent with the experimental results. 
Thus, the nonmonotonic change of the superconducting gap structure with irradiation dose can be semi-quantitatively understood by the increase of Fermi-surface volume and impurity scattering.

Figures\,\ref{gap}(f)-(j) show the momentum direction dependence of the gap function calculated from the effective ${\bf k} \cdot {\bf p}$ model and the estimated $\Delta_{p\rm {,max}} / \Delta_s$ value for each sample. For 3.13\,C/cm$^2$, the ratio $\Delta_{p\rm {,max}} / \Delta_s$ is quite close to 1, and thus the gap near the nodes has relatively weak momentum dependence [Fig.\,\ref{gap}(g)], leading to significant low-energy excitations, which is consistent with the large slope of the quasilinear $T$ dependence of $\Delta\lambda$ [Fig.\,\ref{gap}(b)]. 
The corresponding three-dimensional superconducting gap structures in the pristine and irradiated samples are illustrated in Figs.\,\ref{gap}(k)-(o). We find that in LuPdBi, line nodes surround the [111] and its equivalent corners of the inner FS, in contrast to the previous suggestion for YPtBi~\cite{Kim2018}. Since the position of the line nodes affects the topological nature of LuPdBi, the detection of the nodal positions deserves further experimental investigations.

Finally, we comment on other possibilities. It has been proposed~\cite{Sim2019} that in the $d$-wave dominant pairing state, the $s$-wave pairing can be induced by the broken particle-hole symmetry in the normal state, leading to a nodal superconducting gap. In this $d+s$ state, the chemical potential shift from the charge neutral point is expected to enhance the $s$-wave pairing, leading to a more isotropic gap structure, which is inconsistent with our observation. 
The mixing of $s$-wave singlet and $d$-wave quintet pairing has also been studied theoretically in different types of electronic structures~\cite{Yu2018}. In this singlet-quintet model, it has been shown that the presence of electron pockets along the $\Gamma$-L direction is an important ingredient to realize a gap structure with line nodes. In related materials LuPtBi~\cite{Hou2015} and YbPtBi~\cite{Guo2018}, the nonlinear field dependence of Hall resistivity suggests the presence of such electron pockets. However, in the present LuPdBi, the Hall resistivity shows a perfect $H$-linear behavior (Appendix D) and the band-structure calculations do not show energy dispersions that can induce electron pockets. Thus, although the proposed singlet-quintet pairing state may be relevant for other half-Heusler superconductors, this state is unlikely to be realized in LuPdBi having only the simple hole FSs without electron pockets.


To summarize, from $H_{\rm c2} (T)$ and $\Delta \lambda (T)$ measurements on the pristine and electron-irradiated samples of LuPdBi with $j=3/2$ fermions, we have observed the multigap behaviors and the nonmonotonic change in the gap structure. Considering the simple FSs in LuPdBi, the multigap nature can be regarded as a signature of the strong ASOI and mixed-parity superconducting state. Furthermore, the nonmonotonic evolution of the gap structure with irradiation can be semiquantitatively explained by the simple model based on the mixed-parity state, where the size of the anisotropic $p$-wave septet component of the gap function is enhanced by the carrier doping, and reduced by the impurity scattering effect. These experimental results provide evidence for the exotic mixed-parity singlet-septet state in LuPdBi and demonstrate that electron irradiation can tune the ratio of even- and odd-parity components and even change the nodal topology of the superconducting gap. 

It has been proposed that the existence or absence of gap nodes in the singlet-septet state critically affects novel phenomena, such as topological Majorana surface states~\cite{Timm2017,Wang2017} with intriguing multipole responses~\cite{Kobayashi2019}. Thus our findings provide a novel perspective on research for topological superconducting states.

\bigskip
\noindent
\section*{Acknowledgements}
We thank D. F. Agterberg, P. M. R. Brydon, and E.-G. Moon for fruitful discussions, and K. Ishida, K. Iso, K. Matsuura and T. Yamauchi for technical support. Part of this work was carried out under the Visiting Researcher's Program of the Institute for Solid State Physics, University of Tokyo. We also thank O. Cavani for help in irradiation experiments. Irradiation realized on the SIRIUS platform was supported by the French National network of accelerators for irradiation and analysis of molecules and materials EMIR\&A under Project No. 17-353.

This work was supported by Grants-in-Aid for Scientific Research (KAKENHI) (No.\ JP21J10737, No.\ JP19H00649, No.\ JP18H05227, No.\ JP20H02600, No.\ JP20K21139, No.\ JP19K22123, No.\ JP18H01853, No.\ JP18KK0375, No.\ JP19H01848, and No.\ JP19K21842), Grants-in-Aid for Scientific Research on innovative areas ``Quantum Liquid Crystals" (No.\ JP19H05824, and No.\ JP19H05825), Grant-in-Aid for Scientific Research for Transformative Research Areas (A) ``Condensed Conjugation'' (No.\ JP20H05869) from Japan Society for the Promotion of Science (JSPS), and by CREST (No.\ JPMJCR19T5) from Japan Science and Technology (JST), as well as by the National Science Centre of Poland (No.\ 2015/18/A/ST3/00057) and the Foundation for Polish Science (program START 66.2020).

\renewcommand{\theequation}{A\arabic{equation}}

\section*{Appendix}

\subsection{ELECTRONIC STRUCTURE NEAR THE FERMI LEVEL}

\begin{figure}[b]
    \includegraphics[width=\linewidth]{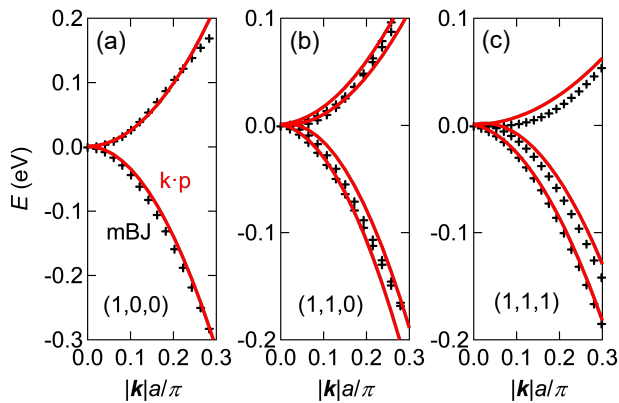}
    \caption{{{\it Ab initio} calculations and effective ${\bf k} \cdot {\bf p}$ model at low energies.} First-principle results of energy dispersions (black crosses) along the (a) $\Gamma$-X, (b) $\Gamma$-K, and (c) $\Gamma$-L directions calculated for LuPdBi with a modified Becke-Johnson (mBJ) potential are fitted with the effective ${\bf k} \cdot {\bf p}$ model (red lines).}
    \label{figS1}
\end{figure}

The calculated band structure of LuPdBi is fitted by the least-squared method using the effective ${\bf k} \cdot {\bf p}$ Hamiltonian represented by Eq.\,(\ref{kp_model}). The four dispersions of the $\Gamma_8$ band along $\Gamma$-X, $\Gamma$-K, and $\Gamma$-L lines are fitted simultaneously, and the fitting range is $|{\bf k}| < 0.3 \pi/a$. The optimized parameters are $\alpha = 3.19$\,eV$a^2/\pi^2$, $\beta = -2.97$\,eV$a^2/\pi^2$, $\gamma = -1.21$\,eV$a^2/\pi^2$, and $\delta = 0.07$\,eV$a/\pi$. Figure\,\ref{figS1} shows theoretical dispersions obtained by {\it ab initio} calculations (black crosses) and the fitting curves (red lines). For more precise fitting, higher-order terms allowed by the crystal symmetry should be considered~\cite{Chadov2010}. However, this effective ${\bf k} \cdot {\bf p}$ Hamiltonian captures the salient features of the electronic structure including the splitting of the Fermi surface for certain directions and the Fermi surface shape, which are important for the superconducting gap structure. 

\subsection{CRYSTAL CHARACTERIZATION}

\begin{table}[t]
    \caption{Chemical compositions of LuPdBi crystals. We give ratios of each constituent element in the pristine, 3.13 (and then irradiated to 5.28\,C/cm$^2$), and 8.44\,C/cm$^2$ (and then irradiated to 11.61 C/cm$^2$) irradiated crystals, obtained by the EDX measurements. }

    \centering
    \begin{tabular}{cccc}
     \hline \hline
    \qquad \qquad \qquad &  Pristine  & \qquad 3.13 (5.28)\,C/cm$^2$ \qquad & \qquad 8.44\,C/cm$^2$ \qquad \\
     \hline 
    Lu  & 33.19\% &32.10\% &32.07\%\\
    Pd & 32.87\% &33.08\% &32.93\%\\
    Bi & 33.94\% &34.82\% &35.00\%\\
    \hline \hline
    \end{tabular}
\end{table}

\begin{figure}[b]
    \includegraphics[width=\linewidth]{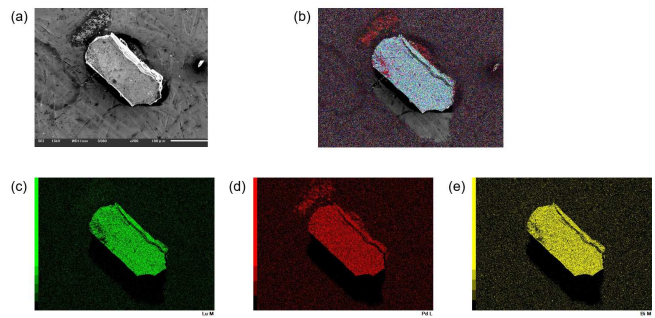}
    \caption{{Distributions of Lu, Pd, and Bi elements.} (a) Scanning electron microscopy (SEM) image of the crystal that is irradiated to 3.13 and then to the 5.28\,C/cm$^2$ dose, taken after the penetration depth measurements. (b) Overlay of EDX analysis of Lu (green), Pd (red), and Bi (yellow) distributions on the SEM image. (c)-(e) Distribution of each (c) Lu, (d) Pd, and (e) Bi element. }
    \label{figS5}
\end{figure}

\begin{figure}
    \centering
    \includegraphics[width=0.8\linewidth]{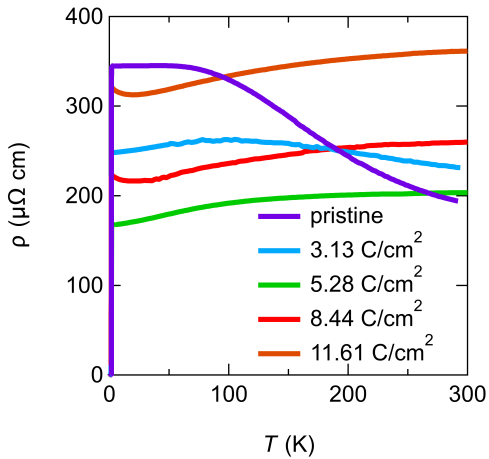}
    \caption{{Temperature dependence of resistivity in LuPdBi before and after irradiation.} 
    Zero-field resistivity is shown as a function of temperature in a wide temperature range for pristine (purple), 3.13 (light-blue), 5.28 (green), 8.44 (red), and 11.61\,C/cm$^2$ (brown) irradiated samples.}
    \label{figS2}
\end{figure}

\begin{figure*}
\includegraphics[width=\linewidth]{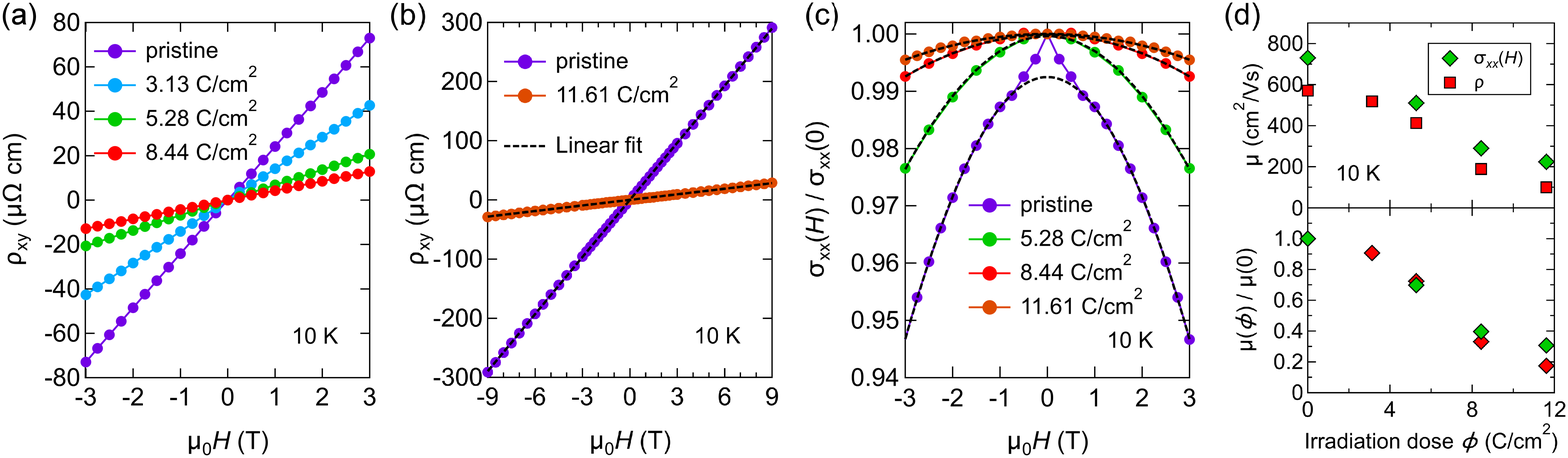}
\caption{{Magnetotransport in LuPdBi before and after irradiation.} (a), (b) Hall resistivity as a function of magnetic field measured at 10\,K in pristine and electron-irradiated samples up to 3\,T (a) and 9\,T (b). (c) Field dependence of the longitudinal magnetoconductivity normalized by the value at zero field measured at 10\,K. The black dashed lines are fitting curves represented as $\sigma_{xx} (H) \propto 1/[1+\mu^2 (\mu_0 H)^2]$. (d) Mobility calculated from the $\sigma_{xx} (H)$ (green diamonds) and the resistivity $\rho$ (red diamonds) as a function of the irradiation dose $\phi$ (upper panel). Normalized mobility by the value in the pristine sample is also shown (lower panel).}
\label{figS3}
\end{figure*}

Obtained crystals were characterized by powder x-ray diffraction and single-crystal x-ray diffraction using an X'pert Pro PANanalytical diffractometer with Cu-K$\alpha$ radiation, and an Oxford Diffraction Xcalibur four-circle diffractometer equipped with Mo-K$\alpha$ radiation, respectively. 
We have checked the homogeneity of our crystals by energy-dispersive x-ray (EDX) spectroscopy as shown in Fig.\,\ref{figS5} and Table I. We find that Lu, Pd, and Bi elements are uniformly distributed in the crystal and the obtained ratios of these constituents are nearly identical. This rules out the possibility that the observed double transition comes from the substantial contamination of $\alpha$-PdBi$_2$ having similar $T_{\rm c}=1.7$\,K with LuPdBi. We also note that the upper critical field of $\alpha$-PdBi$_2$ is only 0.3\,T, which is much smaller than the observed $H_{\rm c2}$.

\subsection{TEMPERATURE DEPENDENCE OF RESISTIVITY}

Figure\,\ref{figS2} shows the temperature dependence of resistivity $\rho(T)$ in the pristine and irradiated samples measured at zero field. In the pristine sample, the resistivity increases as $T$ decreases at high temperatures, $T> 50$\,K, and the resistivity is nearly $T$ independent from 50\,K to $T_c$. The high-temperature semiconducting temperature dependence of resistivity is consistent with the previous reports of half-Heusler materials with low carrier densities~\cite{Nakajima2015,Bay2014}. After electron irradiation, the resistivity shows more metallic $T$ dependence. This is in sharp contrast to the case of ordinary metals with much higher carrier densities, in which electron irradiation causes the upward parallel shift of $\rho(T)$ with increased residual resistivity~\cite{Mizukami2014,Cho2018}. This change in the $T$ dependence of the resistivity is an indication that carriers are introduced by irradiation, which is known for low-carrier systems such as Bi$_2$Te$_3$~\cite{Zhao2016}. Comparisons of 5.28 (green), 8.44 (red), and 11.61\,C/cm$^2$ (brown) data in Fig.\,\ref{figS2} imply that the impurity scattering is also enhanced with an increasing irradiation dose.

\subsection{ESTIMATION OF CARRIER DENSITY AND MOBILITY}

To estimate the carrier density and mobility of the pristine and irradiated samples of LuPdBi, we perform magnetotransport measurements. The magnetic field is applied perpendicular to the current, and the longitudinal ($\rho_{xx}$) and Hall ($\rho_{xy}$) resistivity components are extracted from the symmetric and antisymmetric parts of the raw data against the magnetic field, respectively.  

The Hall resistivity $\rho_{xy}$ shows linear field dependence for all the pristine and irradiated samples, as shown in Figs.\,\ref{figS3}(a) and (b). The carrier density $n_{\rm H}$ is estimated from the Hall resistivity through the equation $d\rho_{xy}/d(\mu_0 H) = 1/(e n_{\rm H})$. The obtained result of $n_{\rm H}$ as a function of the irradiation dose $\phi$ is shown in Fig.\,\ref{irradiation}(a), demonstrating that the hole carriers are introduced by electron irradiation.

The mobility $\mu$ can be obtained from the longitudinal conductivity $\sigma_{xx}$ and resistivity $\rho$. Figure\,\ref{figS3}(c) shows the field dependence of longitudinal magnetoconductivity $\sigma_{xx}(H)/\sigma_{xx}(0)$ obtained from the relation $\sigma_{xx} = \rho_{xx}/(\rho_{xx}^2+\rho_{xy}^2)$. 
From the $\sigma_{xx} (H)$ data, the mobility $\mu$ in a low-carrier system can be calculated through the equation, $\sigma_{xx} (H) \propto 1/[1+\mu^2 (\mu_0 H)^2]$. The black dotted lines in Fig.\,\ref{figS3}(c) represent the fitting curves, showing fairly good agreement with the experiment except for low-field data for the pristine sample. The linear $H$ dependence of magnetoconductivity at low fields for the pristine sample is consistent with the previous studies, and it has been attributed to the weak antilocalization (WAL) effect in the topological surface state~\cite{Pavlosiuk_LuPdBi}. The $H$-linear negative magnetoconductivity at low fields gets smaller as the irradiation dose increases. This evolution of the WAL effect may stem from the $E_F$ shift induced by the electron irradiation, which makes the contribution from the topological surface state to the transport phenomena relatively smaller in the irradiated samples. This smaller weight of the surface state in the irradiated samples is also suggested from the $\rho (T)$ data, which show more metallic behavior in the irradiated samples (Fig.\,\ref{figS2}).

The obtained $\mu$ values from $\sigma_{xx}$ are plotted as the green diamonds in Fig.\,\ref{figS3}(d), which decrease with the irradiation dose $\phi$, indicating that the irradiation introduces impurity scattering as well. The mobility $\mu$ can also be evaluated from the resistivity through the simple relation, $\mu = 1/en_{\rm H} \rho$, and the results are shown as the red diamonds in Fig.\,\ref{figS3}(d). The two independent methods give consistent results of $\mu(\phi)$, although a slight difference in the magnitude of $\mu$ between these two may originate from the WAL effect in the topological surface state mentioned above or quantitative errors in the sample dimensions which affect the absolute value of $\rho$. We note that this quantitative difference is not essential for the analyses of the superconducting gap structure. For the estimation of $\Delta_{p,{\rm max}}$ from Eq.\,(\ref{Delta_p}), we use $\mu(\phi)/\mu(0)$ values calculated from $\sigma_{xx}$ in the pristine, 5.28, 8.44, and 11.61\,C/cm$^2$ irradiated samples, and from $\rho$ in the 3.13\,C/cm$^2$ irradiated sample.

\subsection{UPPER CRITICAL FIELD MEASUREMENTS}

\begin{figure*}
    \includegraphics[width=0.8\linewidth]{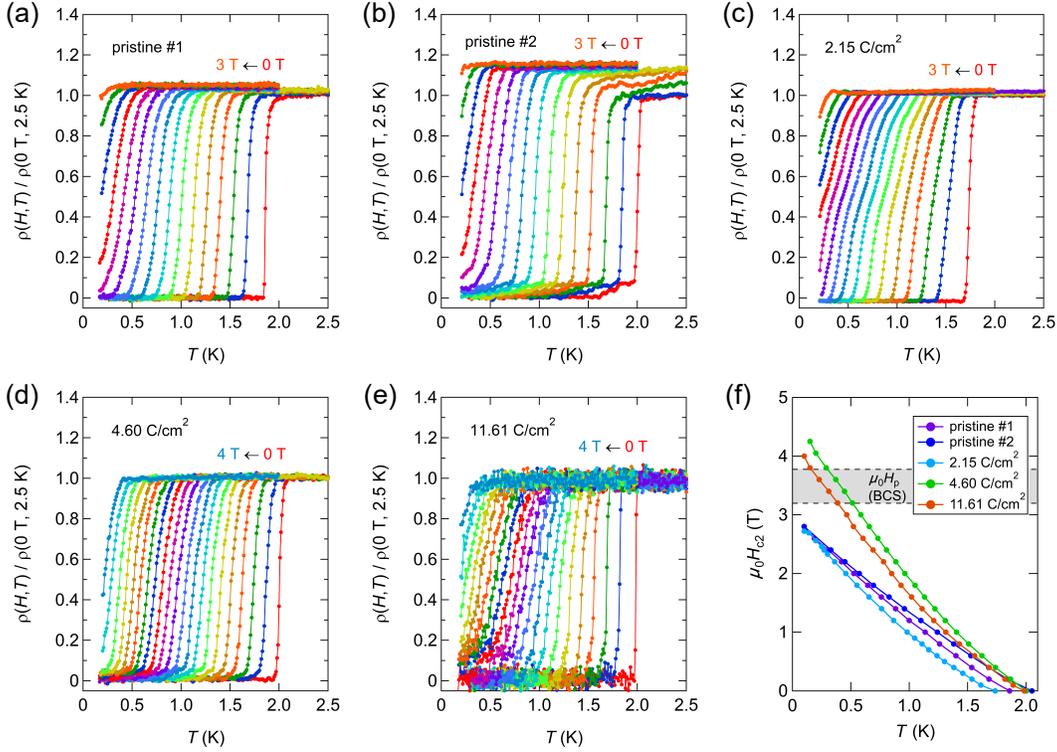}
    \caption{{Temperature dependence of upper critical field determined from resistive transitions in LuPdBi before and after irradiation.} (a)-(e) Temperature dependence of the resistivity measured under various applied magnetic fields, normalized by the zero-field value at 2.5\,K in (a) pristine \#1, (b) pristine \#2, (c) 2.15, (d) 4.60, and (e) 11.61\,C/cm$^2$ samples. The data are plotted for different fields with 0.2-T steps. (f) Temperature dependence of the upper critical field $H_{\rm c2}$ in pristine and irradiated samples. The Pauli limits estimated in BCS theory, $\mu_0 H_{\rm P}\ {(\rm Tesla)} = 1.85$\,$T_{\rm c}\ {(\rm Kelvin)}$ of all measured samples are within the shaded region.}
    \label{figS4}
\end{figure*}

The upper critical fields of the pristine and irradiated samples of LuPdBi are measured by the resistive transitions. The temperature dependence of the resistivity under various fields for pristine and irradiated samples shows apparent shifts of the superconducting transition with field [Figs.\,\ref{figS4}(a)-(e)]. The field-dependent $T_{\rm c}$ is determined as the temperature where the resistivity becomes half of the value at 2.5\,K. The obtained temperature dependence of the upper critical field $H_{\rm c2}$ is summarized in Fig.\,\ref{figS4}(f). We find that at low temperatures $\mu_0 H_{\rm c2} (0)$ reaches high values above 4\,T in 4.60 and 11.61\,C/cm$^2$ irradiated samples. These values exceed the simple estimate of the BCS Pauli limit, $\mu_0 H_{\rm P} {(\rm Tesla)} = 1.85$\,$T_{\rm c} {(\rm Kelvin)}$ (see discussions in Sec. III.B).

Another noticeable feature is that when we compare the $H_{\rm c2}$ data for the pristine \#2, 4.60, and 11.61\,C/cm$^2$ irradiated samples with almost identical zero-field $T_{\rm c}$ values, the initial slope $d H_{\rm c2} / d T |_{T_{\rm c}}$ shows nonmonotonic dose dependence; $d H_{\rm c2} / d T |_{T_{\rm c}}$ is the steepest for the modestly irradiated (4.60\,C/cm$^2$) sample. In general, impurity scattering tends to enhance the initial slope, because the effective coherence length $\xi$ decreases as the mean free path $\ell$ becomes shorter. In the present LuPdBi case, however, this simple impurity scattering effect cannot explain the nonmonotonic behavior of $d H_{\rm c2} / dT |_{T_{\rm c}}$. Other factors that can affect the initial slope of $H_{\rm c2}(T)$ are the Fermi velocity $v_{\rm F}$ and superconducting gap size $\Delta$ because the coherence length in the BCS limit is represented as $\xi = \hbar v_{\rm F} / (\pi\Delta)$. The carrier doping effect of the electron irradiation [Fig.\,\ref{structure}(c)] implies that $v_{\rm F}$ related to the $E({\bf k})$ dispersion curvature increases with the irradiation dose. Furthermore, from the superconducting gap structure analysis in the main text, the anisotropic superconducting gap component in the 11.61\,C/cm$^2$ irradiated sample is expected to be smaller than that in the 4.60\,C/cm$^2$ irradiated sample. Thus, the combination of changes in the Fermi level and superconducting gap induced by irradiation can qualitatively account for the nonmonotonic behavior of the initial slope $d H_{\rm c2} / d T |_{T_{\rm c}}$. 

Finally, the mean free path $\ell=v_{\rm F}\tau$ calculated from the mobility and carrier density of the pristine and 5.28\,C/cm$^2$ samples are nearly equal ($\sim 46$\,nm). This can be understood by the compensation of the carrier doping effect that increases the Fermi velocity $v_{\rm F}$ and the impurity scattering effect that increases the scattering rate $1/\tau$. The mean free path drops to $\sim 31$\,nm and $\sim 27$\,nm in the 8.44 and 11.61\,C/cm$^2$ irradiated samples, respectively, suggesting that the impurity scattering effect dominates at these high irradiation doses. On the other hand, the coherence length estimated from the initial slope of the upper critical field has little change against irradiation (between 11 and 14\,nm). This ensures the condition that the coherence length is still shorter than the mean free path.







\begin{thebibliography}{99}

\bibitem{Chadov2010}
S. Chadov {\it et al}., 
{\it Tunable Multifunctional Topological Insulators in Ternary Heusler Compounds,}
{Nat. Mater.} 
{\bf 9}, 541 (2010).

\bibitem{Lin2010}
H. Lin  {\it et al}., 
{\it Half-Heusler Ternary Compounds as New Multifunctional Experimental Platforms for Topological Quantum Phenomena,}
{Nat. Mater.} 
{\bf 9}, 546 (2010).

\bibitem{Xiao2010}
D. Xiao, Y. Yao, W. Feng, J. Wen, W. Zhu, X. Q. Chen, G. M. Stocks, Z. Zhang, 
{\it Half-Heusler Compounds as a New Class of Three-Dimensional Topological Insulators,}
{Phys. Rev. Lett.} 
{\bf 105}, 096404 (2010).

\bibitem{Liu2016}
Z.\,K. Liu {\it et al}., 
{\it Observation of Unusual Topological Surface States in Half-Heusler Compounds LnPtBi (Ln=Lu, Y),}
{Nat. Commun.} 
{\bf 7}, 12924 (2016).

\bibitem{Suzuki2016}
T. Suzuki {\it et al}., 
{\it Large Anomalous Hall Effect in a Half-Heusler Antiferromagnet,}
{Nat. Phys.} 
{\bf 12}, 1119 (2016).


\bibitem{Hirschberger2016}
M. Hirschberger {\it et al}., 
{\it The Chiral Anomaly and Thermopower of Weyl Fermions in the Half-Heusler GdPtBi,}
{Nat. Mater.} 
{\bf 15}, 1161 (2016).

\bibitem{Kumar2018}
N. Kumar, S.\,N. Guin, C. Felser, and C. Shekhar, 
{\it Planar Hall Effect in the Weyl Semimetal GdPtBi,}
{Phys. Rev. B} 
{\bf 98}, 041103(R) (2018).


\bibitem{Butch2011}
N. P. Butch, P. Syers, K. Kirshenbaum, A. P. Hope, and J. Paglione, 
{\it Superconductivity in the Topological Semimetal YPtBi,}
{Phys. Rev. B} 
{\bf 84}, 220504(R) (2011).

\bibitem{Pavlosiuk_LuPdBi}
O. Pavlosiuk, D. Kaczorowski, and P. Wi{\' s}niewski,
{\it Shubnikov - de Haas Oscillations, Weak Antilocalization Effect and Large Linear Magnetoresistance in the Putative Topological Superconductor LuPdBi,}
{Sci. Rep.} 
{\bf 5}, 9158 (2015).

\bibitem{Nakajima2015}
Y. Nakajima {\it et al}., 
{\it Topological {\it R}PdBi Half-Heusler Semimetals: A New Family of Noncentrosymmetric Magnetic Superconductors,}
{Sci. Adv.} 
{\bf 1}, e1500242 (2015).

\bibitem{Meinert2016}
M. Meinert, 
{\it Unconventional Superconductivity in YPtBi and Related Topological Semimetals,}
{Phys. Rev. Lett.} 
{\bf 116}, 137001 (2016).

\bibitem{Kim2018}
H. Kim {\it et al}., 
{\it Beyond Triplet: Unconventional Superconductivity in a Spin-3/2 Topological Semimetal,}
{Sci. Adv.} 
{\bf 4}, eaao4513 (2018).

\bibitem{Gor'kov2001}
L. P. Gor'kov and E. I. Rashba, 
{\it Superconducting 2D System with Lifted Spin Degeneracy: Mixed Singlet-Triplet State,}
{Phys. Rev. Lett.} 
{\bf 87}, 037004 (2001).

\bibitem{Frigeri2004}
P. A. Frigeri, D. F. Agterberg, A. Koga, and M. Sigrist, 
{\it Superconductivity without Inversion Symmetry: MnSi versus CePt$_{3}$Si,} 
{Phys. Rev. Lett.} 
{\bf 92}, 097001 (2004).

\bibitem{Bauer2012}
E. Bauer and M. Sigrist, 
{\it Non-centrosymmetric Superconductors: Introduction and Overview,} 
Vol. 847 (Springer, Berlin, Heidelberg 2012).

\bibitem{Yuan2006} 
H. Q. Yuan, D. F. Agterberg, N. Hayashi, P. Badica, D. Vandervelde, K. Togano, M. Sigrist, and M. B. Salamon, 
{\it $S$-wave Spin-Triplet Order in Superconductors without Inversion Symmetry: Li$_2$Pd$_3$B and Li$_2$Pt$_3$B,}
Phys. Rev. Lett. {\bf 97}, 017006 (2006).

\bibitem{Nishiyama2007} 
M. Nishiyama, Y. Inada, and G.-q. Zheng, 
{\it Spin Triplet Superconducting State due to Broken Inversion Symmetry in Li$_2$Pt$_3$B,}
Phys. Rev. Lett. {\bf 98}, 047002 (2007).

\bibitem{Brydon2016}
P. M. R. Brydon, L. Wang, M. Weinert, and D. F. Agterberg,
{\it Pairing of $j=3/2$ Fermions in Half-Heusler Superconductors,}
{Phys. Rev. Lett.} 
{\bf 116}, 177001 (2016).

\bibitem{Dutta2021}
P. Dutta, F. Parhizgae, and A. M. Black-Schaffer,
{\it Superconductivity in Spin-3/2 Systems: Symmetry Classification, Odd-Frequency Pairs, and Bogoliubov Fermi Surfaces,}
{Phys. Rev. Research} 
{\bf 3}, 033255 (2021).

\bibitem{Kim_arxiv}
H. Kim {\it et al}., 
{\it Anomalous Quantum Oscillations in Spin-3/2 Topological Semimetal YPtBi,}
arXiv:2010.12085 (2010). 

\bibitem{Boettcher2018}
I. Boettcher and  I. F. Herbut, 
{\it Unconventional Superconductivity in Luttinger Semimetals: Theory of Complex Tensor Order and the Emergence of the Uniaxial Nematic State,}
{Phys. Rev. Lett.} 
{\bf 120}, 057002 (2018).

\bibitem{Venderbos2018}
J. W. F. Venderbos, L. Savary, J. Ruhman, P. A. Lee, and L. Fu, 
{\it Pairing States of Spin-$\frac{3}{2}$ Fermions: Symmetry-Enforced Topological Gap Functions,}
{Phys. Rev. X} 
{\bf 8}, 011029 (2018).

\bibitem{Yu2018}
J. Yu, and C.-X. Liu, 
{\it Singlet-Quintet Mixing in Spin-Orbit Coupled Superconductors with $j = \frac{3}{2}$ Fermions,}
{Phys. Rev. B} 
{\bf 98}, 104514 (2018).

\bibitem{Sim2019}
G. B. Sim, A. Mishra, M. J. Park, Y. B. Kim, G. Y. Cho, S. B. Lee, 
{\it Topological $d+s$ Wave Superconductors in a Multiorbital Quadratic Band Touching System,}
{Phys. Rev. B} 
{\bf 100}, 064509 (2019).

\bibitem{Roy2019}
B. Roy, S. A. A. Ghorashi, M. S. Foster, and A. H.  Nevidomskyy,
{\it Topological Superconductivity of Spin-$3/2$ Carriers in a Three-Dimensional Doped Luttinger Semimetal,}
{Phys. Rev. B} 
{\bf 99}, 054505 (2019).

\bibitem{Timm2017}
C. Timm, A. P. Schnyder, D. F. Agterberg, and P. M. R. Brydon, 
{\it Inflated Nodes and Surface States in Superconducting Half-Heusler Compounds,}
{Phys. Rev. B} 
{\bf 96}, 094526 (2017).

\bibitem{Wang2017}
W. Yang, T. Xiang, and C. Wu, 
{\it Majorana Surface Modes of Nodal Topological Pairings in Spin-$\frac{3}{2}$ Semimetals,}
{Phys. Rev. B} 
{\bf 96}, 144514 (2017).

\bibitem{Kobayashi2019}
S. Kobayashi, A. Yamakage, Y. Tanaka, and M. Sato, 
{\it Majorana Multipole Response of Topological Superconductors,}
{Phys. Rev. Lett.} 
{\bf 123}, 097002 (2019).


\bibitem{Billington1961}
D. S. Billington and J. H. Crawford, 
{\it Radiation Damage in Solids}
(Princeton University Press, Princeton, NJ, 1961).

\bibitem{Zhao2016}
L. Zhao {\it et al.}, 
{\it Stable Topological Insulators Achieved Using High Energy Electron Beams,}
{Nat. Commun.} 
{\bf 7}, 10957 (2016).

\bibitem{Mizukami2014}
Y. Mizukami {\it et al.}, 
{\it Disorder-Induced Topological Change of the Superconducting Gap Structure in Iron Pnictides,}
{Nat. Commun.} 
{\bf 5}, 5657 (2014).

\bibitem{Cho2016}
K. Cho {\it et al}., 
{\it Energy Gap Evolution across the Superconductivity Dome in Single Crystals of (Ba$_{1-x}$K$_x$)Fe$_2$As$_2$,}
{ Sci. Adv.} 
{\bf 2}, e1600807 (2016).

\bibitem{Takenaka2017}
T. Takenaka {\it et al}., 
{\it Full-Gap Superconductivity Robust against Disorder in Heavy-Fermion CeCu$_{2}$Si$_{2}$,}
{Phys. Rev. Lett.} 
{\bf 119}, 077001 (2017).

\bibitem{Cho2018}
K. Cho, M. Ko{\'{n}}czykowski, S. Teknowijoyo, M. A. Tanatar and R. Prozorov, 
{\it Using Electron Irradiation to Probe Iron-Based Superconductors,}
{Supercond. Sci. Technol.} 
{\bf 31}, 064002 (2018).

\bibitem{Mizukami2017}
Y. Mizukami {\it et al}., 
{\it Impact of Disorder on the Superconducting Phase Diagram in BaFe$_2$(As$_{1-x}$P$_x$)$_2$,}
{J. Phys. Soc. Jpn.} 
{\bf 86}, 083706 (2017).

\bibitem{Ishihara2018}
K. Ishihara {\it et al}., 
{\it Evidence for $s$-Wave Pairing with Atomic Scale Disorder in the van der Waals Superconductor NaSn$_{2}$As$_{2}$,}
{Phys. Rev. B} 
{\bf 98}, 020503(R) (2018).

\bibitem{WIEN2k}
P. Blaha {\it et al.},
{\it WIEN2k: An APW+lo Program for Calculating the Properties of Solids,}
{J. Chem. Phys.} 
{\bf 152}, 074101 (2020).

\bibitem{Tran2009}
F. Tran and P. Blaha,
{\it Accurate band gaps of semiconductors and insulators with a semilocal exchange-correlation potential,}
{Phys. Rev. Lett.} 
{\bf 102}, 226401 (2009).

\bibitem{Prozorov2000}
R. Prozorov, R. W. Giannetta, A. Carrington, and  F. M. Araujo-Moreira, 
{\it Meissner-London State in Superconductors of Rectangular Cross Section in a Perpendicular Magnetic Field,}
{Phys. Rev. B} 
{\bf 62}, 115 (2000).


\bibitem{Bay2014}
T.\,V. Bay {\it et al}.,
{\it Low Field Magnetic Response of the Non-centrosymmetric Superconductor YPtBi,}
{Solid State Commun.} 
{\bf 183}, 13-17 (2014).

\bibitem{Gurevich2003}
A. Gurevich, 
{\it Enhancement of the Upper Critical Field by Nonmagnetic Impurities in Dirty Two-Gap Superconductors,}
{Phys. Rev. B} 
{\bf 67}, 184515 (2003).

\bibitem{Thomas1996}
F. Thomas {\it et al}.,
{\it Strong Coupling Effects on the Upper Critical Field of the Heavy-Fermion Superconductor UBe$_{13}$,}
{J. Low Temp. Phys.} 
{\bf 102}, 117-132 (1996).

\bibitem{Aoki2019}
D. Aoki, K. Ishida, and J. Flouquet, 
{\it Review of U-based Ferromagnetic Superconductors: Comparison between UGe$_2$, URhGe, and UCoGe,}
{J. Phys. Soc. Jpn.} 
{\bf 88}, 022001 (2019).

\bibitem{Tada2008}
Y. Tada, N. Kawakami, and S. Fujimoto, 
{\it Colossal Enhancement of Upper Critical Fields in Noncentrosymmetric Heavy Fermion Superconductors near Quantum Criticality: CeRhSi$_3$ and CeIrSi$_3$,}
{Phys. Rev. Lett.} 
{\bf 101}, 267006 (2008).

\bibitem{Mazidian2013}
B. Mazidian, J. Quintanilla, A. D. Hillier, and J. F.  Annett,  
{\it Anomalous Thermodynamic Power Laws near Topological Transitions in Nodal Superconductors,}
{Phys. Rev. B} 
{\bf 88}, 224504 (2013).

\bibitem{Mishra2009} 
V. Mishra, G. R. Boyd, S. Graser, T. Maier, P. J. Hirschfeld, and D. J. Scalapino,
{\it Lifting of Nodes by Disorder in Extended-$s$-State Superconductors: Application to Ferropnictides,}
Phys. Rev. B {\bf 79}, 094512 (2009).

\bibitem{Hou2015}
Z. Hou {\it et al}., 
{\it High Electron Mobility and Large Magnetoresistance in the Half-Heusler Semimetal LuPtBi,}
{Phys. Rev. B} 
{\bf 92}, 235134 (2015).


\bibitem{Guo2018}
C. Y. Guo {\it et al}., 
{\it Evidence for Weyl Fermions in a Canonical Heavy-Fermion Semimetal YbPtBi,}
{Nat. Commun.} 
{\bf 9}, 4622 (2018).



\end{thebibliography}
\end{document}